\documentclass{aa}
\usepackage{txfonts}
\usepackage{graphicx}
\usepackage{natbib}
\bibpunct{(}{)}{;}{a}{}{,}

\begin{document}

\title{Oligarchic and giant impact growth of terrestrial planets\\ in the
        presence of gas giant planet migration.}
\author{Martyn J. Fogg \and Richard P. Nelson.}
\institute{Astronomy Unit, Queen Mary, University of London, Mile
End Road, London E1 4NS.}
\date{Received/Accepted}
\abstract{Giant planets found orbiting close to their central stars,
the so-called `hot Jupiters', are thought to have originally formed
in the cooler outer regions of a protoplanetary disk and then to
have migrated inward via tidal interactions with the nebula gas. We
present the results of N--body simulations which examine the effect
such gas giant planet migration has on the formation of terrestrial
planets. The models incorporate a 0.5 Jupiter mass planet undergoing
type II migration through an inner protoplanet--planetesimal disk,
with gas drag included. Each model is initiated with the inner disk
being at successively increased levels of maturity, so that it is
undergoing either oligarchic or giant impact style growth as the gas
giant migrates.

\noindent In all cases, a large fraction of the disk mass survives
the passage of the giant, either by accreting into massive
terrestrial planets shepherded inward of the giant, or by being
scattered into external orbits. Shepherding is favored in younger
disks where there is strong dynamical friction from planetesimals
and gas drag is more influential, whereas scattering dominates in
more mature disks where dissipation is weaker. In each scenario,
sufficient mass is scattered outward to provide for the eventual
accretion of a set of terrestrial planets in external orbits,
including within the system's habitable zone. This scattering,
however, significantly reduces the density of solid material,
indicating that further accretion will occur over very long time
scales. A particularly interesting result is the generation of
massive, short period, terrestrial planets from compacted material
pushed ahead of the giant. These planets are reminiscent of the
short period Neptune-mass planets discovered recently, suggesting
that such `hot Neptunes' could form locally as a by-product of giant
planet migration. }
\titlerunning{Growth of terrestrial planets in the
presence of gas giant migration.}
\authorrunning{M.J. Fogg \& R.P. Nelson}

\maketitle

\section{Introduction.}\label{intro}

Over the last ten years the radial velocity technique has been
successfully employed in the detection of giant planets orbiting
nearby main sequence stars \citep[e.g.][]{mayor,butler,marcy}. To
date, 136 extra-solar planetary systems have been discovered, 14 of
them multiple, leading to a total of 155 giant planets\footnote{Data
from the Extrasolar Planets Encyclopedia at
http://www.obspm.fr/encycl/encycl.html; 20/6/05.}. Even though
radial velocity observations are more sensitive to short period
orbits, a surprising discovery has been the substantial population
of giant planets orbiting close to their central star at distances
$\lesssim$ 0.1 AU. Twenty nine such objects are known (3 of them
sited in multiple systems) comprising $\sim$ 20\% of the total
sample. These so-called `hot Jupiters' are mostly sub-jovian in
mass, with low eccentricity orbits, and are often found associated
with stars more metal rich than the Sun
\citep{udry,santos1,fischer}.

The most likely scenario for the origin of hot Jupiters involves
initial formation further out in the nebula beyond the `snowline' as
per conventional formation theories \citep[e.g.][]{pollack},
followed by an episode of inward orbital migration, propelled by
tidal interactions between the planet and gas disk. A variety of
orbital migration phenomena have been proposed for protoplanets as
they grow in mass, but the one most likely to operate in this case
is type II migration where the planet has become sufficiently
massive ($\ga 100\ \mathrm{M}_\oplus$) to open up a gap in the gas,
thereby migrating inward at a rate controlled by the disk viscous
time scale \citep{lin1,lin2,nelson1}. This time scale is typically
on the order of a few $\times 10^5$ years. What finally stops the
inward drift of these planets at such small radial distances, other
than fortuitous disk dispersal, is presently unknown and it may be
that some migrating planets are accreted by the central star.

It appears therefore that hot Jupiter systems are not uncommon and
possibly represent an extreme rearrangement of planetary mass during
formation as compared with the Solar System. This comparison raises
an interesting set of questions. What effect would hot Jupiter
migration have on terrestrial planet formation? Might accretion of
terrestrial planets be interrupted or prevented altogether? Can we
have any realistic expectation of ever discovering Earth--like
planets in these systems, or is it probable their habitable zones
are empty of significant material, having been swept clean by the
passage of the giant? The assumption that hot Jupiter systems are
barren are among those advanced to support such speculative concepts
as the "Rare Earth" hypothesis \citep{ward1} and the Galactic
Habitable Zone \citep{lineweaver1,lineweaver2}, both of which argue
for a large number of special circumstances required to form an
Earth--like planet. From their standpoint, any significant
difference in planetary system architecture from that of the solar
system is likely to preclude the formation or survival of planets
that are habitable for multi-cellular life.

These issues were examined by \citet{armitage} with a simple
time-dependent disk model of the evolution of gas, dust and
planetesimals. He assumed that the effect of giant planet migration
is to first sweep the inner disk clear of planetesimals, and then
looked at whether a new generation of planetesimals could be formed
in the terrestrial planet region by the subsequent resupply of dust
from outer regions by advection and diffusion. His conclusion was
that resupply of solid material into the inner disk would be
inefficient and terrestrial planet formation would be unlikely. In
contrast, \citet{mandell} consider a late migration scenario and use
N-body simulations to model the migration of a Jupiter mass planet
through a fully formed terrestrial planet system. The typical
pattern of evolution observed included: 1) excitation of planetary
orbits by sweeping resonances with the inward migrating giant; 2)
close encounters between the planets resulting in collisions or
mutual scattering; and 3) slingshot encounters with the giant as it
passed through the inner system, leading to ejection, collision with
the central star, or scattering into bound exterior orbits with
increased semi-major axis, eccentricity and inclination. Short
migration times allowed a larger fraction of planets to survive
rather than being ejected whereas the trend was reversed for long
migration times. Overall, $\sim$ 25\% of the planets survived in a
wide variety of orbits exterior to the giant, orbits which, they
speculated, might subsequently become circularized as a result of
dynamical friction with outer system planetesimals or interaction
with the remnant gas disk. They concluded therefore that inward
migration of a giant planet does not invariably eliminate pre-formed
terrestrial planets and that, given an initial layout of bodies
similar to that of the Solar System, between $\sim$ 1-4\% of systems
in which migration occurred could still possess a planet in the
habitable zone. Actual formation of terrestrial planets in the
presence of a hot Jupiter has been modelled by \citet{raymond}. They
assume a previous rapid migration of the giant into its final close
orbit and then model the later stages of terrestrial planet
accretion from an exterior protoplanet disk using N-body methods.
Their conclusion is that the presence of a hot Jupiter does little
to interfere with terrestrial planet formation outside of an annulus
that is within a factor of three in period to the giant (about a
factor of two in semi-major axis). Planet formation in the habitable
zone is not adversely affected.

The conclusions of these three papers span the widest possible range
of outcome, from the occurrance of terrestrial planets in hot
Jupiter systems being highly unlikely, through possible but rare, to
commonplace. This variation originates from the very different
assumptions and initial conditions in each case. \citet{armitage}
does not model the dynamic effects of the migrating giant on
planetesimals and instead assumes total loss of planetary building
blocks from within the swept zone. This may be unrealistic
especially as, by the time a giant planet has grown large enough to
start type II migration, considerable accretion into larger
planetary embryos could have already occurred in the inner system,
bodies which might not be so readily accreted or bulldozed into the
central star. The picture of \citet{mandell} of a giant migrating
through a mature terrestrial planet system may suffer from
unrealistic timing as giant planet formation and migration is
constrained to occur within the $\sim 10^6 - 10^7$ year lifetime of
the gas disk whereas the terminal `giant impacts' phase of
terrestrial planet formation is thought to last $\sim 10^8$ years
\citep[e.g.][]{chambers2}. The optimistic conclusions of
\citet{raymond} are a function of their initial condition of placing
the hot Jupiter in its final, post migration, orbit and then
modelling terrestrial planet formation in an essentially
undisturbed, unmixed, exterior disk. This requires rapid giant
planet formation and migration followed by the formation of a new
terrestrial disk from the remaining debris which must somehow be
reduced back to a more unevolved, damped, and chemically
differentiated state.

Timing issues are therefore important in the study of this problem,
involving initially isolated sequences of events inside and beyond
the nebula snowline. Whilst there remains much uncertainty over
detail, one fairly certain constraint which gives an upper limit to
the time available is that giant planets must both form and complete
migration in considerably less than $10^7$ years, before the
dispersal of the nebular gas. Observations suggest that 50\% of
stars in clusters have lost their disks by an age of $\sim 3\times
10^6$ years \citep{haisch}. Estimation of a lower age limit is more
problematic as it must rely on our incomplete theories of giant
planet formation. If the core-accretion model is to be preferred,
favorable conditions would allow giants to form in $\gtrsim 10^6$
years \citep{pollack,papaloizou3,alibert}. In the meantime,
accretion will be ongoing within the planetesimal swarm in the
terrestrial planet region. According to the current picture, an
early phase of runaway growth will give way to a more lengthy phase
of oligarchic growth where similar sized protoplanets emerge from
the swarm in well-spaced orbits which remain near circular due to
dynamical friction from the surrounding sea of planetesimals
\citep{kokubo1}. Oligarchic growth ends when planetesimal numbers
decline to the extent that their damping effect on protoplanet
orbits becomes insufficient to prevent orbit crossing. This
inaugurates the last phase of terrestrial planet formation, that of
so-called `giant impacts', involving the mutual accretion of
protoplanets and thinning down of their number to the point where
the final planets emerge, positioned in stable non-crossing orbits.
Simulations of this final stage of terrestrial planet growth suggest
that it would take $\sim 10^8$ years to complete \citep{chambers2},
long after the disappearance of the nebular gas. Oligarchic growth
however starts much earlier, whilst gas is still present:
simulations by \citet{kokubo2} have shown that it takes only $\sim\
5\times10^5$ years to generate $\sim 0.01 - 0.03\ \mathrm{M}_\oplus$
planetary embryos from a planetesimal disk at 1 AU. Thus, in the
case of a giant planet migrating through the terrestrial planet
zone, it seems most probable that this would occur at some time
within, or towards the end of the phase of oligarchic growth in that
region.

In order to better satisfy these timing constraints, we present
below a set of N-body simulations of giant planet migration through
progressively evolved inner system protoplanet-planetesimal disks
undergoing either late oligarchic or early giant impact style
growth. In Section 2 we outline our model and its initial
conditions; in Section 3 the results are presented and discussed; in
Section 4 we consider some caveats and future model improvements,
and in Section 5 we offer our conclusions.

\section{Description of the model.}\label{description}
\subsection{Forces.}\label{forces}
We choose to model planet growth and migration using the
hybrid-symplectic N-body simulation package \emph{Mercury 6}
\citep{chambers1}, modified to include the effect of gas drag on
planetesimals and type II migration on a single giant planet. The
ingredients required for the simulations are thus one migrating
giant planet, and an interior disk of small protoplanets embedded
within a swarm of planetesimals. However, due to the huge number of
particles involved, the realization of a realistic planetesimal
disk, with every body treated as fully interacting, is well beyond
the current state of the art. We proceed therefore by following the
so-called $N + N'$ approach of \citet{thommes2} where we have $N$
protoplanets embedded in a disk of $N'$ `super-planetesimals',
particles that represent an idealized ensemble of a much larger
number of real planetesimals. The giant and the protoplanets feel
all the modeled gravitational forces, whereas the
super-planetesimals feel gravitational forces from the central star,
protoplanets, and giant planet, but are otherwise non
self-interacting. This prevents their relatively high masses from
unrealistically auto-exciting the disk. Super-planetesimals alone
also experience gas drag with the drag being calculated using a
defined physically realistic planetesimal radius. The issue of the
mass of the super-planetesimals was addressed by \citet{thommes2}
who found in test runs that protoplanets undergo effective dynamical
friction if super-planetesimal masses are $\la 0.1$ times the
initial protoplanet masses.

The coordinate origin is based on the central star. The acceleration
experienced by each of the super-planetesimals is given by:
\begin{eqnarray}
\frac{d^2\vec{r}_i}{dt^2} = & - & \frac{GM_\ast\vec{r}_i}{|\vec{r}_i|^3}
- \sum^N_{j=1} \frac{Gm_j\vec{r}_{ij}}{|\vec{r}_{ij}|^3} - \sum^N_{j=1}
\frac{Gm_j\vec{r}_j}{|\vec{r}_{j}|^3} \\ \nonumber
& - & \sum^{N'}_{k=1} \frac{Gm_k\vec{r}_k}{|\vec{r}_{k}|^3} +
\vec{a}_{\mathrm{drag}},
\label{acc_plan}
\end{eqnarray}
where $M_\ast$ is the stellar mass, $m_j$ and $\vec{r}_j$ are
particle masses and position vectors, and
$\vec{r}_{ij}=\vec{r}_i-\vec{r}_j$. The first term on the right hand
side represents the acceleration from the central star,
the second term the accelerations from the protoplanet and giant planet,
the third and fourth terms are the indirect terms arising from
the acceleration of the coordinate system, and
$\vec{a}_{\mathrm{drag}}$  is the acceleration due to gas drag. \\
The acceleration experienced by the protoplanets and gas giant planet is
given by:
\begin{eqnarray}
\frac{d^2\vec{r}_i}{dt^2} = & - & \frac{GM_\ast\vec{r}_i}{|\vec{r}_i|^3}
- \sum^N_{j=1} \frac{Gm_j\vec{r}_{ij}}{|\vec{r}_{ij}|^3}(1- \delta_{ij})
- \sum^{N'}_{k=1} \frac{Gm_k\vec{r}_{ik}}{|\vec{r}_{ik}|^3} \\ \nonumber
& - & \sum^N_{j=1} \frac{Gm_j\vec{r}_j}{|\vec{r}_{j}|^3}
- \sum^{N'}_{k=1} \frac{Gm_k\vec{r}_k}{|\vec{r}_{k}|^3}
+ \vec{a}_{\mathrm{TypeII}}.
\label{acc_proto}
\end{eqnarray}
The first term again represents the acceleration due to the central star,
the second and third terms represent the accelerations due to the
protoplanets/giant planet and super-planetesimals, respectively, and
$\delta_{ij}$ is the Kronecker delta function. The fourth and fifth terms are
indirect terms, and $\vec{a}_{\mathrm{TypeII}}$ is the acceleration
driving type II migration of the gas giant planet alone.

Planetesimals are small enough to experience a drag force from
moving through the nebula gas. This acceleration, which acts both to
cause an inward radial drift and a damping of eccentricities and
inclinations, takes the form:

\begin{equation}\label{drag}
\vec{a}_{\mathrm{drag}}=-\frac{1}{2m_{\mathrm{pl}}} C_{\mathrm{D}}
\pi r^2_{\mathrm{pl}} \rho_{\mathrm{g}} |\vec{u}|\vec{u}\ ,
\end{equation}
where $m_{\mathrm{pl}}$ and $r_{\mathrm{pl}}$ are the physical mass
and radius of a single planetesimal respectively, $C_{\mathrm{D}}$
is the drag coefficient (taken here to be $C_{\mathrm{D}} = 1$) and
$\vec{u}=\vec{v}_{\mathrm{p}} - \vec{v}_{\mathrm{g}}$ is the
velocity of the planetesimal with respect to the gas. The gas is
assumed to move in a circular orbit which, due to pressure support,
revolves at slightly less than Keplerian speed. The relation is:

\begin{equation}\label{}
\vec{v}_{\mathrm{g}} =
\left(1-2\eta\right)^{\frac{1}{2}}\vec{v}_{\mathrm{K}}\ ,
\end{equation}
where $\vec{v}_{\mathrm{K}}$ is the local Keplerian velocity and
$\eta=0.0019(a/1\mbox{AU})^{1/2} $, given the nebula scale height
introduced later in Eq.~\ref{height} \citep{adachi}.

In order to model type II migration of a giant planet, we adopt a
simple prescription and assume that this occurs at a rate controlled
by a local viscous disk evolution timescale that is proportional to
orbital period. Assuming an alpha viscosity model, this is roughly:

\begin{equation}\label{viscous}
\tau_{\nu} = \frac{2}{3}\left(\frac{r}{h}\right)^2\left(\alpha\Omega
\right)^{-1},
\end{equation}
where $r$ is the radial distance, $h$ is the disk scale height,
$\alpha$ is the alpha parameter and $\Omega$ is the orbital
frequency. Given $\alpha = 2 \times 10^{-3}$, $h/r = 0.05$ and $r =
5\ \mbox{AU}$ then $\tau_\nu \approx 0.25\ \mbox{Myr}$. Since
$\tau_\nu \propto a^{1.5}$, where $a$ is the semi-major axis,\
$\dot{a} \propto a^{-0.5}$ so inward migration speeds up as it
proceeds. (Note that this value of $h/r$ differs slightly from that
obtained from Eq.~\ref{height}.)

The type II migration process is also assumed to exert strong
eccentricity and inclination damping which is taken here to operate
over a timescale $\ = \tau_\nu/50$. The acceleration
$\vec{a}_{\mathrm{TypeII}}$ is therefore implemented as:

\begin{equation}\label{}
\vec{a}_{\mathrm{TypeII}} = - \frac{\vec{v}}{2\tau_\nu} - 25 \left[
\frac{2(\vec{v}\cdot\vec{r})\vec{r}}{r^2\tau_\nu}
+\frac{2(\vec{v}\cdot\vec{k})\vec{k}}{\tau_\nu} \right],
\end{equation}
where $\vec{v}$ is the giant planet's velocity vector and
$\vec{k}$ is a unit vector in the vertical direction. Note that the
factor of 2 appearing in the the first term on the right hand side
arises because the migration time is half the angular momentum
removal time.

\begin{figure*}
\centering
 \includegraphics[width=17cm]{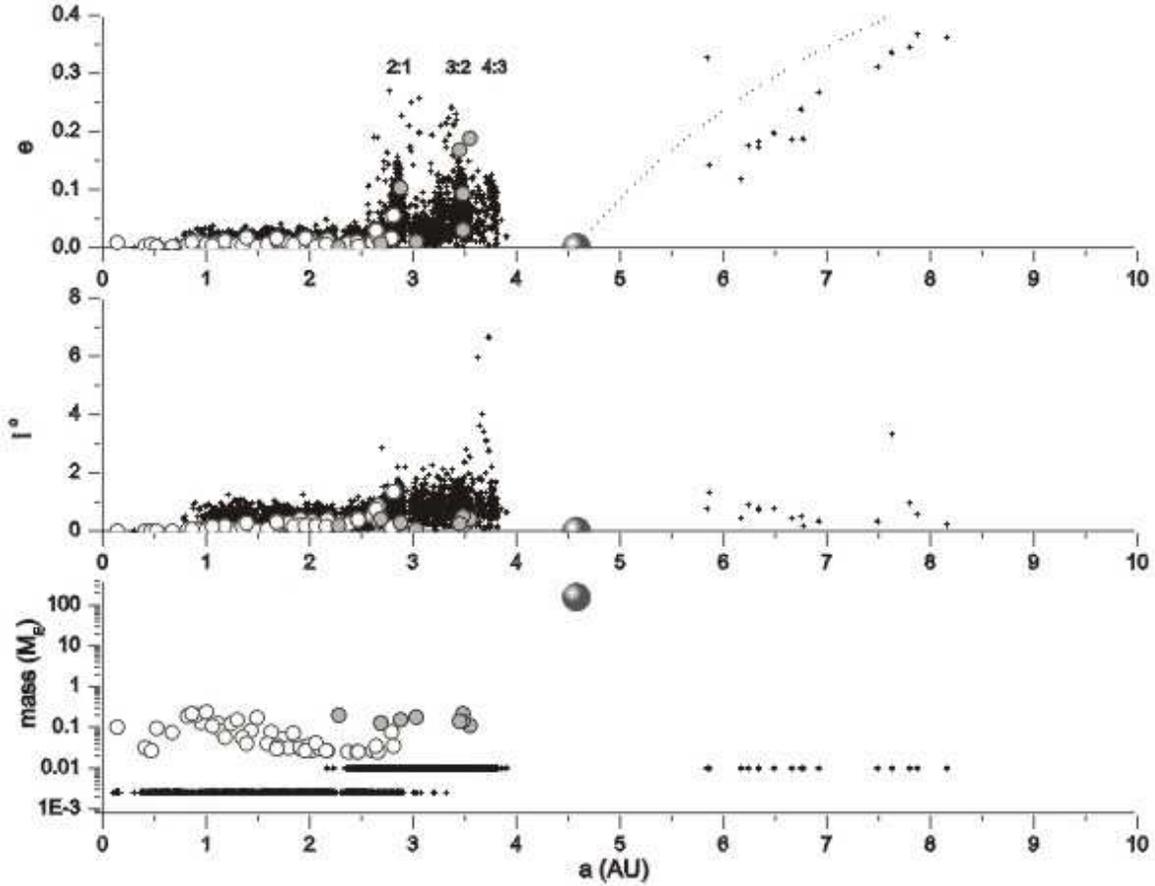}
 \caption{Scenario I at $20\,000$ years after the start of giant planet
 migration, showing the mass, inclination and eccentricity of objects.
 Small black dots represent super-planetesimals; white
 filled circles are rocky protoplanets; grey filled circles are
 icy protoplanets and the large highlighted grey filled circle is
 the giant. The dotted line in the upper panel shows the eccentricity
 at which the pericentre of an exterior object intersects the orbit of
 the giant. The location of the 2:1, 3:2 and 4:3 resonances with the
 giant are indicated.}
 \label{figure:1}
\end{figure*}

\subsection{The Nebula.}\label{nebula}
The nebula model used for this study is based on the minimum mass
solar nebula (MMSN) of \citet{hayashi} and is defined as follows.

The surface density of solids is:

\begin{equation}\label{}
\Sigma_{\mathrm{s}} = f_{\mathrm{neb}} f_{\mathrm{ice}} \Sigma_1
\left( \frac{a}{1\mathrm{AU}}\right)^{-1.5}\ ,
\end{equation}
where $f_{\mathrm{neb}} $ is a nebular mass scaling factor,
$\Sigma_1 = 7\ \mbox{g cm}^{-2} $ and the ice condensation
coefficient $f_{\mathrm{ice}}=1$ for $a< 2.7\ \mbox{AU} $ (the
distance chosen for the nebula `snowline') and
$f_{\mathrm{ice}}=4.2$ for $a \ge 2.7\ \mbox{AU} $.

The volume density of gas is:

\begin{equation}\label{}
\rho_{\mathrm{g}} = f_{\mathrm{neb}} \rho_1
\left(\frac{a}{1\mathrm{AU}}\right)^{-\frac{11}{4}}
\exp\left[-z^2/h^2\right]\ ,
\end{equation}
where $\rho_1 = 2.0\times10^{-9}(f_{\mathrm{gas}}/240)(\Sigma_1/10)\
\mbox{g cm}^{-3} $, $f_{\mathrm{gas}} $ is the gas to dust ratio,
$z$ is the height from the midplane of the nebula and the disk scale
height $h$ is taken to be:

\begin{equation}\label{height}
h = 0.045\left(\frac{a}{\mathrm{AU}}\right)^{\frac{5}{4}} .
\end{equation}

In the context of the core-accretion model of giant planet
formation, a significant additional mass of solids seems to be
required in order to form a core quick enough to initiate gas
accretion before the loss of the gaseous component of the nebula
\citep{lissauer,pollack,thommes2}. Moreover, since hot Jupiters are
usually found around stars more metal-rich than the Sun, a greater
solids content might reasonably be expected within their
protoplanetary nebulae. Thus, for this study we assume an equivalent
mass of $3\times\mbox{MMSN}$ of solids. However, as the nebula is
taken to be somewhat evolved at our starting point, considerable gas
could already have been lost so we assume a smaller equivalent mass
of $2\times\mbox{MMSN}$ of gas. The relevant parameters are
therefore set to $f_{\mathrm{neb}} = 3$ and $f_{\mathrm{gas}} =
160$.

\begin{figure*}
\centering
 \includegraphics[width=17cm]{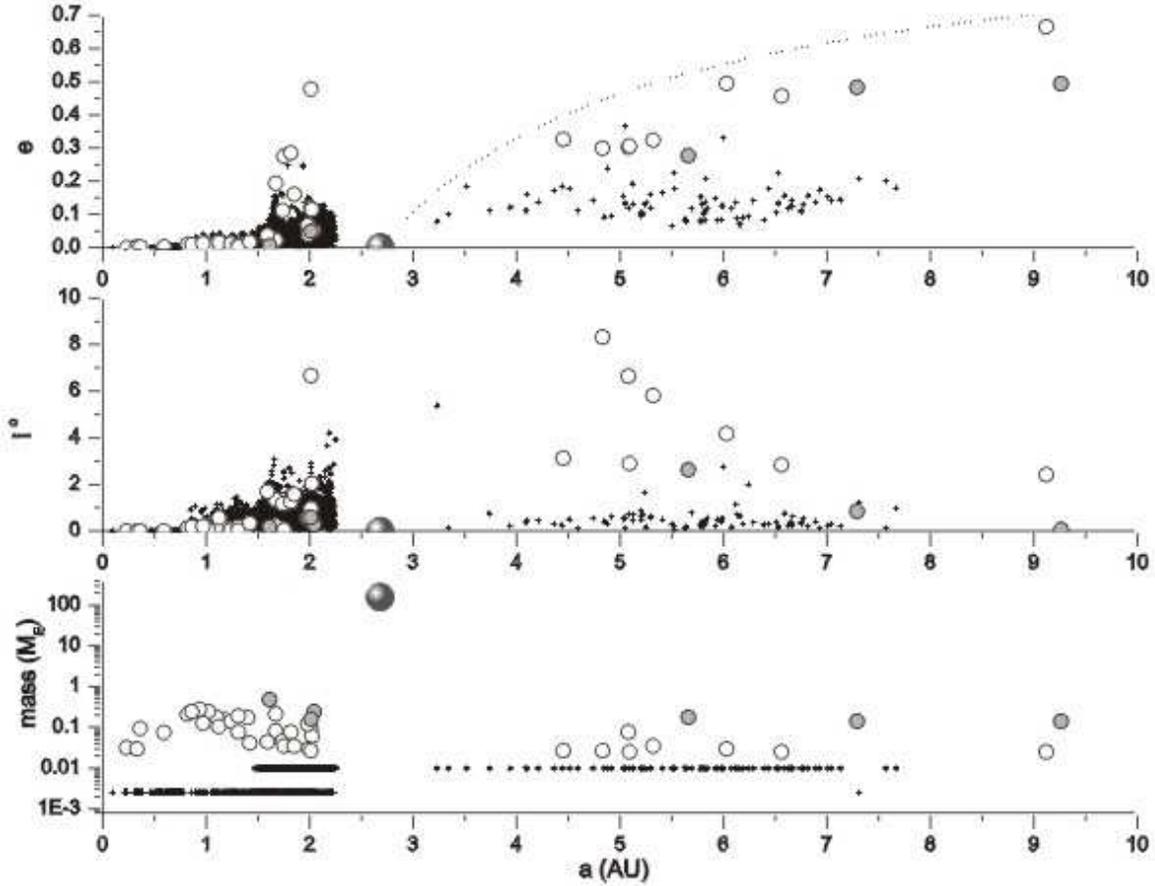}
 \caption{Scenario I at $100\,000$ years after the start of giant planet
 migration. The giant has now moved inward to 2.68 AU and has scattered
 a significant amount of mass into exterior orbits.}
 \label{figure:2}
\end{figure*}

\begin{table}
\caption{Initial disk set-up} %
\label{table:1}  %
\centering
\begin{tabular}{c| c c| c}
 \hline\hline %
& Rocky Zone & Icy Zone& Total\\
& 0.4--2.7~AU & 2.7--4.0~AU & 0.4--4.0~AU\\
 \hline
$M_{\mathrm{solid}}$ & $9.99~\mathrm{M}_{\oplus}$ &
$14.8~\mathrm{M}_{\oplus}$ & $24.8~\mathrm{M}_{\oplus}$\\ %
 \hline
$m_{\mathrm{proto}}$ & $0.025~\mathrm{M}_{\oplus}$ &
$0.1~\mathrm{M}_{\oplus}$\\ %
$N$ & 66 & 9 & 75\\ %
 \hline
$m_{\mathrm{s-pl}}$ & $0.0025~\mathrm{M}_{\oplus}$ &
$0.01~\mathrm{M}_{\oplus}$\\ %
$N'$ & 3336 & 1392 & 4278\\
 \hline
$f_{\mathrm{proto}}$ & 0.17 & 0.06 & 0.1\\ %
 \hline\hline
\end{tabular}
\end{table}

\subsection{Initial conditions and running of the
simulations.}\label{initcond} The interior disk of protoplanets and
planetesimals is modeled initially from 0.4--4 AU which, given the
nebula parameters already described in Section~\ref{nebula}, amounts
to $M_{\mathrm{solid}} = 24.8~\mathrm{M}_{\oplus}$ of solid (rocky
or icy) material. Adopting the oligarchic growth picture advanced by
\citet{kokubo2}, we set a nominal age for the disk of 0.5 Myr, by
which point protoplanets in the inner disk may have grown to a few
percent of an Earth mass and perhaps to greater than this beyond the
snowline. Initial protoplanet masses of $m_{\mathrm{proto}} = 0.025\
\mathrm{M}_{\oplus}$ and $m_{\mathrm{proto}} = 0.1\
\mathrm{M}_{\oplus}$ are therefore chosen to represent bodies
interior and exterior to the snowline respectively. The number $N$
of protoplanets was calculated based on an assumption that the
average radial spacing between them is 8 mutual Hill radii.
Semi-major axes for this number of protoplanets are then generated
randomly with probabilities weighted in order to reproduce the disk
surface density profile. Eccentricities and inclinations are
randomized from a Rayleigh distribution with RMS values of 0.01 and
0.005 respectively. Additional orbital elements required are
randomized uniformly from within their range. The total mass of
protoplanets is then subtracted from the total mass of the disk
annulus and the mass that remains is divided into $N'$
super-planetesimals with a mass ($m_{\mathrm{s-pl}}$) one tenth that
of the protoplanets in their zone. The orbital elements of the
super-planetesimals are then generated in the same manner. Note that
the super-planetesimal mass $m_{\mathrm{s-pl}}$ is distinct from the
physical planetesimal mass $m_{\mathrm{pl}}$ which is used solely
for the purpose of calculating gas drag (see Eq.~\ref{drag}).

\begin{figure*}
\centering
 \includegraphics[width=17cm]{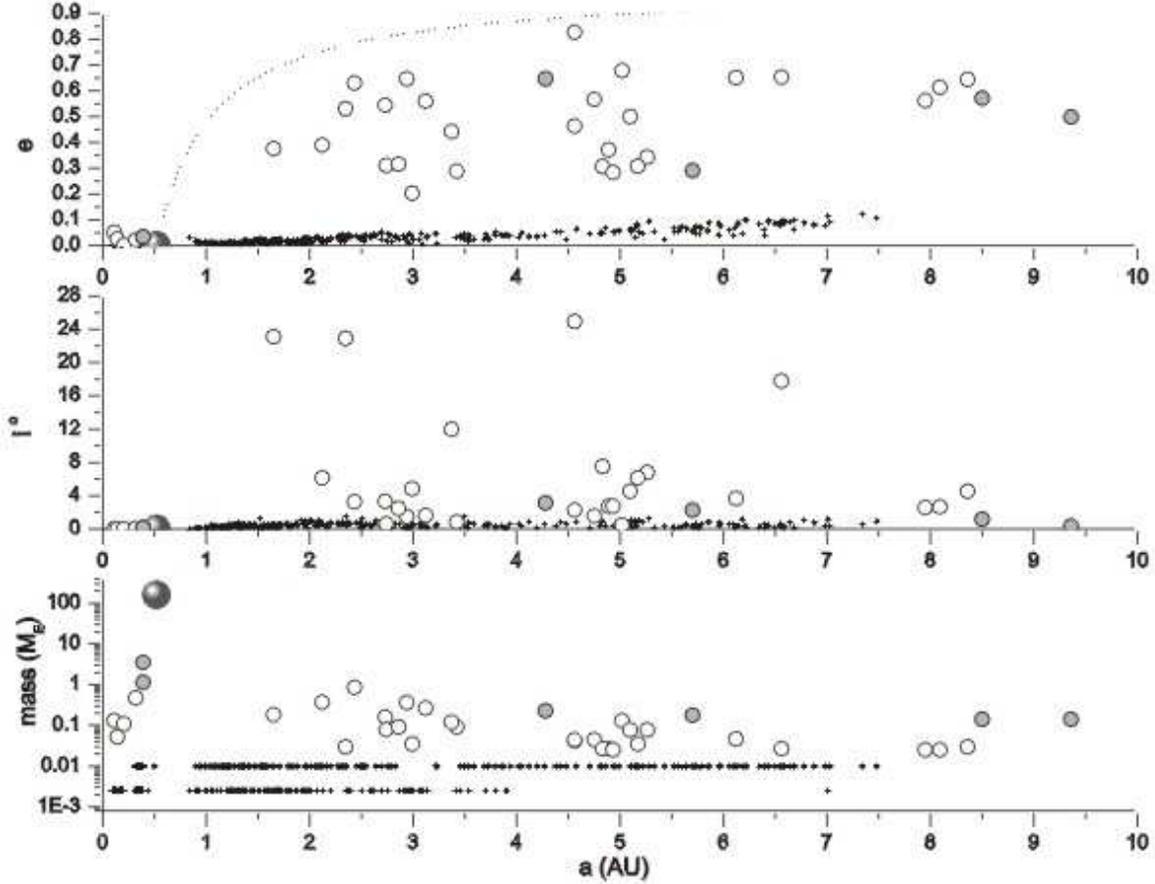}
 \caption{Scenario I at $160\,000$ years after the start of giant planet
 migration. The giant has now moved inward to 0.52 AU. The scattered
 exterior disk has grown, but a substantial amount of mass in planetesimals
 and six rapidly accreting protoplanets remain interior to the giant. The
 outer three of these six are in first order resonances with the giant.}
 \label{figure:3}
\end{figure*}

Data for this initial disk model are shown in Table \ref{table:1}.
The overall values are $N = 75$ and $N' = 4278$ and the mass
fraction of the disk contained in protoplanets $f_{\mathrm{proto}}
\approx 0.1$. The parameter $f_{\mathrm{proto}}$ is used here as a
rough measure of the evolution of the disk and we take
$f_{\mathrm{proto}} = 0.5$, the point where the total mass in
protoplanets exceeds that in planetesimals, to denote the transition
between oligarchic and giant impact growth regimes.

\emph{Mercury 6} models accretion as inelastic collisions between
objects whose radii are calculated from an input density, values of
1, 2 and 3 $\mbox{g~cm}^{-3}$ for giant, icy and rocky masses
respectively being chosen here. All collision permutations are
enabled, except that between super-planetesimals, and a physically
realistic planetesimal radius of $r_{\mathrm{pl}} = 10~\mathrm{km}$
is imposed both for calculating drag and collisions. The mass of the
central star is taken to be $M_\ast = 1~\mathrm{M}_\odot$.

\begin{table}
\caption{Overall disk data: after 0.1--3.0 Myr of
evolution} %
\label{table:2}  %
\centering  %
\begin{tabular}{c| c c c c c}
 \hline\hline
Time~(Myr) & 0.1 & 0.25 & 0.5 & 1.0 & 3.0\\
Scenario ID & I & II & III & IV & V\\
 \hline
$M_{\mathrm{solid}}~(\mathrm{M}_{\oplus})$ & 24.4 & 23.7 & 22.9 & 20.6 & 14.6\\
$m_{\mathrm{max}}~(\mathrm{M}_{\oplus})$ & 0.22 & 0.33 & 0.52 & 0.91 & 1.28\\
$N$ & 56 & 51 & 40 & 29 & 19\\
$N'$ & 3863 & 3312 & 2660 & 1341 & 499\\
$f_{\mathrm{proto}}$ & 0.20 & 0.24 & 0.32 & 0.51 & 0.70\\ %
 \hline\hline
\end{tabular}
\end{table}

Before the giant is added to the picture, five instances of this
disk were allowed to evolve by being run for 0.1, 0.25, 0.5, 1.0 and
3.0~Myr, the purpose being to provide the basis for five type II
migration scenarios through progressively evolved inner system
material. In these runs, the timestep for the symplectic part of
Mercury's hybrid integrator was set to 8 days and all mass straying
inward of 0.1~AU was eliminated and added to the mass of the central
star. Data for these evolved disks are given in Table \ref{table:2}.
It can be seen that the amount of mass lost to the central star,
principally via gas drag-induced orbital decay of planetesimals,
remains modest (except in the case of Scenario V) but that as the
disk ages the maximum protoplanet mass $m_{\mathrm{max}}$ increases
and particle numbers decrease. As would be expected,
$f_{\mathrm{proto}}$ increases with time until by 1~Myr
$f_{\mathrm{proto}} > 0.5$ and giant impact style growth has begun.

The five type II migration scenarios studied here are constructed
from these five evolved disks. A giant planet of mass
$0.5~\mathrm{M_J}$ is placed into each simulation at 5~AU and
allowed to migrate inward with $\tau_{\nu}(5~\mathrm{AU}) =
237\,000\ \mathrm{years}$ (see Eq.~\ref{viscous}). The giant
migrates down to 0.1 AU in $\ga 160\,000$ years, the exact time
depending on the amount of mass remaining interior to its orbit. All
simulations therefore were run for $170\,000$ years, with the type
II migration algorithm being switched off once the giant reached 0.1
AU. In order to better model processes when the giant migrates down
to small radial distances, collision with the central star is
computed when $r < 0.014\ \mathrm{AU} \cong 3~\mathrm{R}_{\odot}$,
the approximate radius of a T-Tauri star. The initial timestep
chosen for the symplectic integrator was 1 day; but as dynamical
spreading and the effects of migration and drag drives some material
into closer orbits, the timestep was reduced as the simulation
progressed. The position of the inner edge of the swarm was
therefore monitored at intervals during each run keeping the
timestep close to one tenth the orbital period of the innermost
object. This considerably increased the run times of these
simulations, especially those based on younger disks, requiring
about a month of processor time for completion.

\begin{figure}
 \resizebox{\hsize}{!}{\includegraphics{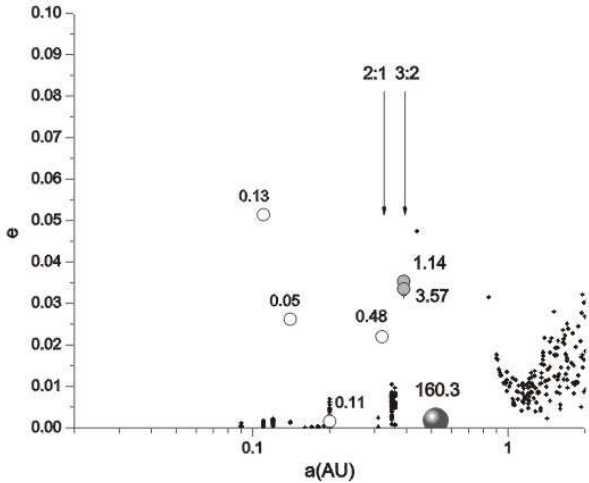}}
 \caption{Detail of the interior regions of Scenario I at $160\,000$ years
 after the start of giant planet migration, showing $e$ vs $a$ for
 objects $\leq 2$~AU. Planetary masses are indicated in units of
 $\mathrm{M}_\oplus$. A total of $\sim 15 \mathrm{M}_\oplus$ of material
 has been pushed inward by the giant, $\sim$ two thirds of it
 remaining in super-planetesimals, many of which do not show
 individually in this diagram as they are over-plotted in
 the vicinity of the 3:2 and 2:1 resonances.}
 \label{figure:4}
\end{figure}

\section{Results.}\label{results}
\subsection{Typical features of a run.}\label{features}
The character of the planetary systems formed from these runs was
found to vary systematically with the age of the inner disk.
However, all scenarios also exhibited a number of behavioral
features in common. We discuss these first by describing one of the
scenarios in detail.

Five snapshots of the evolution of Scenario I are illustrated in
Figs.~1--5 showing the mass, inclination and eccentricity of objects
vs. semi-major axis. The original provenance of the protoplanets
(interior or exterior to the snowline) is denoted by the shading of
its symbol as described in the caption to Fig.~\ref{figure:1}. In
the case of a merger between rocky and icy protoplanets, this
shading is determined by that of the most massive of the pair.

A juncture early in the simulation is illustrated in
Fig.~\ref{figure:1}, 20\,000 years after the start of migration when
the giant has moved inward to 4.58 AU. Several prominent features
have developed. The outer edge of the disk has been pushed inward at
the 4:3 resonance with the giant -- a shepherding process that acts
primarily on planetesimals that are damped by gas drag
\citep{tanaka}. In addition, sweeping resonances have captured a
substantial population of planetesimals and some protoplanets at 3:2
and 2:1, increasing their concentration at these locations and
exciting their orbits. The overall effect is a compaction and
excitation of the outermost annulus of the disk between the 4:3 and
2:1 resonances. Inwards of this zone, the presence of the
approaching giant has had little influence.

The state of play some time later at 100\,000 years is shown in
Fig.~\ref{figure:2}. The giant has now moved inward to 2.68 AU,
continuing to push the outer edge of the disk ahead of it at the 4:3
resonance. An increased amount of mass has been entrained in the
region between 4:3 and 2:1 and resonant pumping and mutual
scattering has raised the eccentricities of some protoplanets to
high values. This has allowed some objects to cross the gap between
the disk and giant whereupon a close slingshot encounter causes
expulsion into an exterior orbit. A diffuse and excited exterior
disk is now in the process of formation, composed predominantly of
the more weakly damped protoplanet material.

At 160\,000 years the giant has moved inward to 0.52 AU as shown in
Fig.~\ref{figure:3}. The most prominent feature now appears to be
the scattered exterior disk made up of numerous protoplanets with
large $a$ and $e$ and a diffuse population of planetesimals with a
$\Sigma_\mathrm{s}$ of only a few percent of the original disk. This
however still represents the minority of solids mass. Two thirds of
the disk mass remains interior to the giant in the form of remaining
planetesimals and six protoplanets. A blow-up of the interior
regions of the system at this juncture is shown in
Fig.~\ref{figure:4}. A total of $\sim 15 \mathrm{M}_\oplus$ of solid
material has been compacted interior to the giant, most of which
lies between 0.32--0.39~AU. Two massive protoplanets of 3.57 and
1.14~$\mathrm{M}_\oplus$ are captured at the 3:2 resonance with the
giant and a smaller 0.48~$\mathrm{M}_\oplus$ protoplanet is found at
the 2:1 resonance. The majority of the mass however remains in
super-planetesimals entrained at these resonances (and mostly
over-plotted in the figure) and in a ring of matter between them.
Protoplanetary eccentricities therefore remain low, even at
resonances, due to strong dynamical friction. In addition, accretion
rates onto these objects have now become so high that collisional
damping is also acting to control the growth of their
eccentricities.

\begin{figure*}
\centering
 \includegraphics[width=17cm]{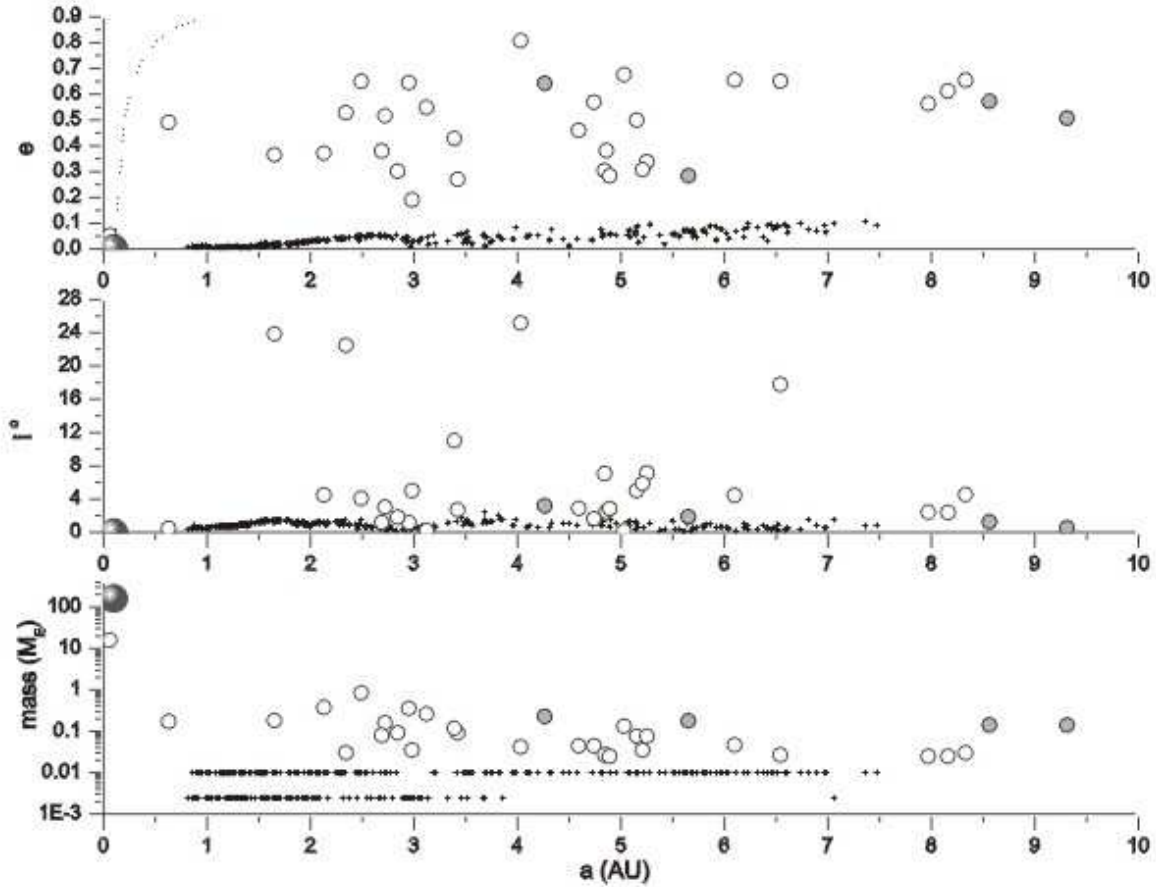}
 \caption{Scenario I at $170\,000$ years after the start of giant planet
 migration. The giant has now stopped at 0.1 AU. All interior mass has
 accreted into a single $\sim~16~\mathrm{M}_\oplus$ object after an intense
 episode of runaway accretion and giant impacts.}
 \label{figure:5}
\end{figure*}

The final snapshot of Scenario I at 170\,000 years is shown in
Fig.~\ref{figure:5}. The giant has stopped its migration at 0.1 AU
and, whilst the character of the exterior disk is unchanged,
evolution has proceeded rapidly to $f_{\mathrm{proto}} = 1$ within
the compacted interior material. No planetesimals remain, having
either been swept up by protoplanets or accreted by the central
star. An intense episode of runaway accretion and giant impacts has
ended in the assembly of a single $15.65~\mathrm{M}_{\oplus}$ planet
out of resonance with the giant at 0.055 AU. This arrangement of
close orbiting giant and an inner Neptune-mass planet has a striking
similarity to the two innermost planets in the \object{55~Cancri}
system \citep{mcarthur}.

\begin{figure*}
\centering
 \includegraphics[width=17cm]{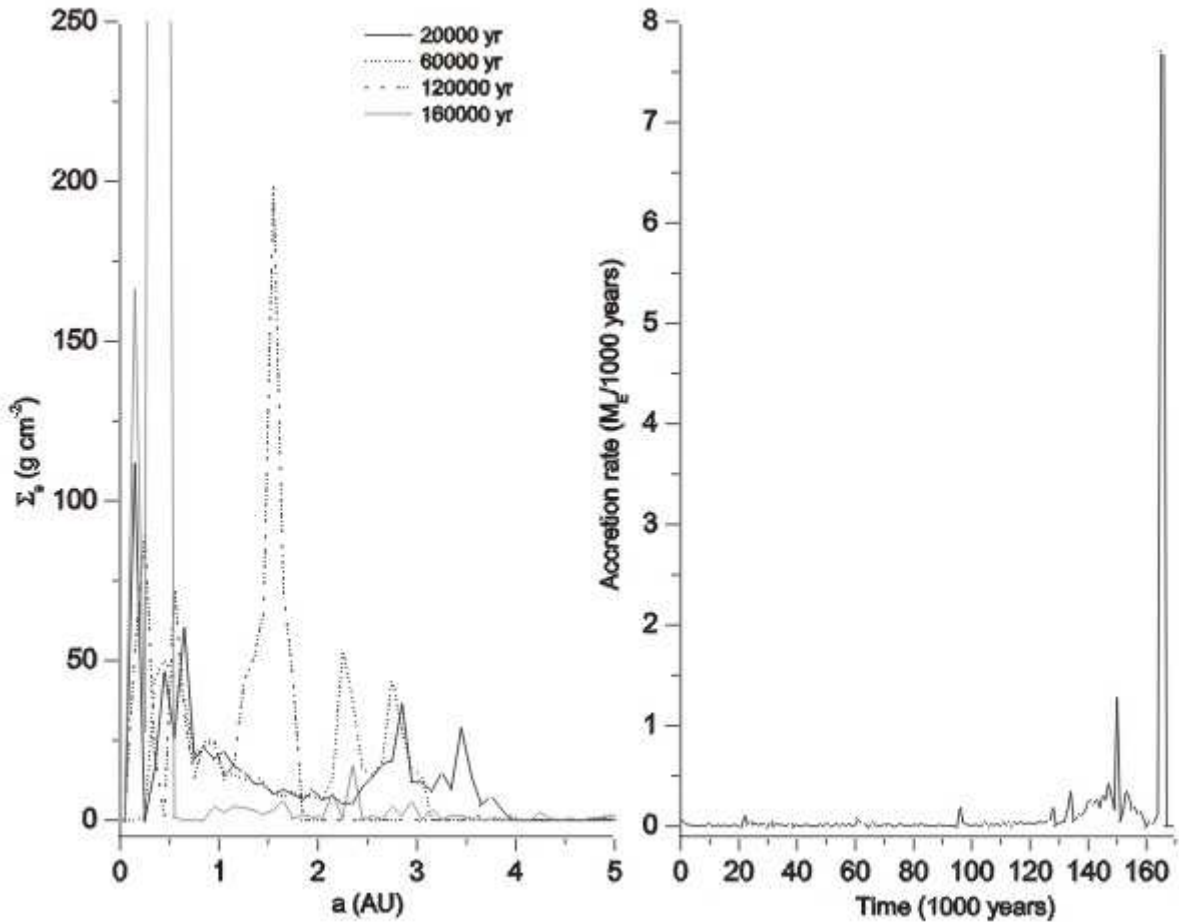}
 \caption{Surface density evolution (left hand panel) and accretion rates
 (right hand panel) for Scenario I. Growing surface density peaks at the 2:1
 and 3:2 resonances sweep through the inner system ahead of the giant.
 Accretion rates increase after $\sim 120\,000$ years and the final intense
 accretion spike represents clear up of remaining material shepherded within
 0.1 AU.}
 \label{figure:6}
\end{figure*}

To reinforce the above interpretation of processes at work in
Scenario I, the surface density evolution of the disk and its
accretion rate are shown in Fig.~\ref{figure:6}. The left hand panel
shows the disk surface density profile (obtained by summing all
protoplanets and super-planetesimals in 0.1~AU width bins) at
20\,000, 60\,000, 120\,000 and 160\,000 years after the start of
migration; the right hand panel plots the amount of mass accreted
onto protoplanets only (including giant impacts) every 1\,000 years
for the duration of the run. In the $\Sigma_{\mathrm{s}}$ plot, two
surface density enhancements are clearly visible as spikes at the
3:2 and 2:1 resonances and are seen to grow whilst moving inward. At
120\,000 years, these have merged into one: the outer half of the
original disk having by now been squeezed into a dense ring. By
160\,000 years, most of this mass is now confined within 0.5 AU and
$\Sigma_{\mathrm{s}}$ here has risen to
$\sim~10^3~\mathrm{g~cm}^{-2}$ (an increase by a factor of $\ga$~10
over the previous, undisturbed, disk surface density) which is off
the vertical scale in the figure. The effect of this disk compaction
process is seen clearly in the accretion rate plot. Mass accretion
rises significantly after 120\,000 years due to both the resultant
high values of $\Sigma_{\mathrm{s}}$ and the fact that much of this
mass now resides in a zone where dynamical times are shorter. Growth
interior to the giant ends in a short-lived and dramatic phase of
runaway planetesimal accumulation and giant impacts within the
material pushed into the small volume inside 0.1 AU from the central
star.

The behavioral features seen to a greater or lesser extent in all
the runs summarize as follows.

\begin{enumerate}
 \item \emph{Shepherding of planetesimals.} As planetesimal random
 velocities are continuously damped by gas drag, their tendency is
 to be pushed inward, ahead of the giant. Shepherding of
 protoplanets also results as a weaker secondary effect as they are,
 to a varying extent, coupled to the planetesimal disk by dynamical
 friction.
 \item \emph{Resonant capture.} First order resonances with the
 giant gather an increasing amount of mass as they sweep inward.
 This, in addition to the shepherding effect eventually results in
 the compacting of some of the disk mass into a zone close to the
 central star.
 \item \emph{Acceleration of planetary growth interior to the
 giant.} Accretion speeds up within the compacted interior disk.
 This is particularly rapid within the disk remnant squeezed inside
 0.1 AU, where the final evolutionary phases of accretion are rushed
 through in mere thousands rather than millions of years. Typically,
 1--3 massive close orbiting planets are the end result. Where there
 is one survivor, its mass and configuration can be reminiscent of
 the `hot Neptune' type of planet identified recently.
 \item \emph{Creation of a scattered exterior disk.} Pumping of
 eccentricities at resonances and by mutual perturbations permit
 some disk material to undergo a close encounter with the giant
 where it is scattered into an external orbit. Protoplanets, being
 less strongly damped, are more likely to have this happen than
 planetesimals. The result is a dynamically excited and widely
 dispersed external disk of material where accretion rates are
 greatly reduced. Leaving aside the issue of interaction of this
 ejecta with the disk of gas and solids outside the giant's original
 formation orbit
 (not simulated here), it seems likely from the result shown in
 Fig.~\ref{figure:5} that further giant impact style evolution
 ending with a set of planets in non-crossing orbits would take much
 longer than the $\sim 10^8$ years estimated for the solar system.
\end{enumerate}

\subsection{Dependence on the maturity of the inner
disk.}\label{maturity}
The purpose of running five scenarios through
a progressively more mature inner disk is to explore the issue of
whether the timing of migration has any systematic effect on the
results. This is possible because when a disk evolves and small
objects accumulate onto larger ones, both dynamical friction and gas
drag become less effective overall, influencing both the shepherding
and scattering behaviors previously described.

\begin{table*}
\caption{Fate of the disk mass at 170\,000 years} %
\label{table:3}  %
\centering  %
\begin{tabular}{c| c c c c c}
 \hline\hline %
Scenario & I & II & III & IV & V\\
 \hline\hline %
Total Initial Solids $(\mathrm{M}_{\oplus})$ & 24.39 & 23.68 & 22.89
& 20.61 & 14.59\\
 \hline %
Total Surviving Solids $(\mathrm{M}_{\oplus})$ & 22.06~(90\%) &
21.04~(89\%) & 12.05~(53\%) & 14.75~(70\%) & 9.91~(68\%)\\
 \hline %
Interior Surviving Solids $(\mathrm{M}_{\oplus})$ & 15.65~(64\%) &
14.60~(62\%) & 4.96~(22\%) & 2.86~(14\%) & 0.00~(0\%)\\
$N, f_{\mathrm{proto}}$ & 1, 1 & 3, 1 & 1, 0.99 & 1, 1 & 0, 0\\
 \hline %
Exterior Surviving Solids $(\mathrm{M}_{\oplus})$ & 6.41~(26\%) &
6.44~(27\%) & 7.09~(31\%) & 11.59~(56\%) & 9.91~(68\%)\\
$N, f_{\mathrm{proto}}$ & 32, 0.66 & 32, 0.63 & 25, 0.65 & 23, 0.93
& 16, 0.89\\
 \hline %
Accreted by Star $(\mathrm{M}_{\oplus})$ & 1.05~(4\%) & 0.74~(3\%) &
9.60~(42\%) & 5.81~(28\%) & 3.03~(21\%)\\
 \hline %
Accreted by Giant $(\mathrm{M}_{\oplus})$ & 1.19~(5\%) & 1.88~(8\%)
& 1.24~(5\%) & 0.34~(2\%) & 1.65~(11\%)\\
 \hline %
Ejected $(\mathrm{M}_{\oplus})$ & 0.00~(0\%) & 0.025~(0.1\%) &
0.00~(0\%) & 0.025~(0.1\%) & 0.00~(0\%)\\
 \hline\hline %
\end{tabular}
\end{table*}

\begin{figure}
 \resizebox{\hsize}{!}{\includegraphics{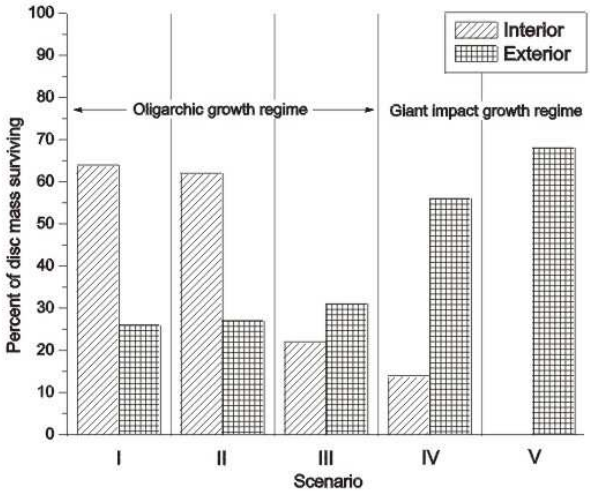}}
 \caption{Interior and exterior surviving solids as a percentage of
 initial disk mass at 170\,000 years.}
 \label{figure:7}
\end{figure}

In all the simulations there were five possible fates awaiting all
the modeled disk particles: 1) survival in a body orbiting interior
to the giant; 2) survival in a body orbiting exterior to the giant;
3) accretion by the central star; 4) accretion by the giant; and 5)
ejection from the system. These data are shown in Table
\ref{table:3} which lists the fate of the disk mass at the end of
the simulations: the total surviving solids and the five end points
being shown as a percentage of the total initial solids. It is
noticeable from Table \ref{table:3} that systematic trends in the
fractionation of solids between some of these end points do appear.

The most obvious trends are that the percentage of the original disk
mass that ends up surviving interior to the giant falls with disk
maturity whereas that expelled into exterior orbits rises with disk
maturity (see Fig.\ref{figure:7}). The protoplanet mass fraction
$f_{\mathrm{proto}} = 1$ for the inner material is indicative of its
rapid evolution, whereas the values of $f_{\mathrm{proto}} \approx$
0.6 -- 0.9 for the outer material are as much influenced by the
preferential tendency for protoplanets to be widely scattered as it
is by their previous growth. The implication is that scattering is
noticeably more effective in older, less dissipative disks. Giant
planet migration through a protoplanet/planetesimal disk rapidly
advances its evolution (as measured by $f_{\mathrm{proto}}$), both
by speeding up accretion and by fractionating objects according to
the magnitude of their damping, but disks in an oligarchic stage of
growth have more of a tendency to push material ahead of the giant
and disks undergoing giant impact type growth allow a greater amount
of mass to escape into external orbits.

The amount of mass accreted by the giant or ejected are minor in all
cases but the picture is complicated by the statistics of mass
accreted by the central star. In the last three scenarios, this is
significant. However, the high values in Scenarios III \& IV result
mainly from the impact of a single massive protoplanet during the
final energetic phase of accretion within 0.1~AU. The loss of mass
to the star in Scenario V resulted from the orbital decay of a
substantial annulus of super-planetesimals in the absence of any
remaining interior protoplanet which could accrete them. The
stochastic fate of large individual bodies at late times can
therefore overwrite and partially obscure systematic trends in the
data. In Scenario III, for example, if a single close encounter near
the end of the simulation had resulted in a giant impact rather than
a scattering into the star, a single interior planet of ($\sim 12.5\
\mathrm{M}_\oplus$) would have formed instead -- another hot Neptune
analogue.

\begin{table*}
\caption{Interior surviving planets at 170\,000 years compared with
the four currently known `hot Neptune' systems. Data include the
closest giant planet.} %
\label{table:4}  %
\centering  %
\begin{tabular}{c| c c c c c}
 \hline\hline %
Scenario & ID & Mass~($\mathrm{M}_{\oplus}$) & a~(AU) & e & Resonances\\
 \hline\hline
I & OLI8 & 15.65 & 0.055 & 0.052 & None\\
& GIA1 & 160.3 & 0.100 & 0.003 & \\
 \hline
II & OLI9 & 2.33 & 0.047 & 0.148 & 5:4 with ICE9, 5:3 with ICE1\\
& ICE9 & 6.69 & 0.054 & 0.071 & 4:3 with ICE1\\
& ICE1 & 5.58 & 0.065 & 0.022 & \\
& GIA1 & 160.9 & 0.100 & 0.005 & \\
 \hline
III & ICE1 & 4.89 & 0.023 & 0.110 & None\\
& GIA1 & 160.3 & 0.100 & 0.001 & \\
 \hline
IV & OLI28 & 2.86 & 0.036 & 0.035 & None\\
& GIA1 & 159.8 & 0.100 & 0.002 & \\
 \hline\hline %
\object{GJ~436} & b & 21.3 $\sin i$ & 0.028 & 0.12 & None\\
 \hline
\object{55~Cancri} & e & 14.3 $\sin i$ & 0.038 & 0.174 & None\\
& b & 267.3 $\sin i$ & 0.11 & 0.02 & \\
 \hline
\object{$\mu$~Arae} & d & 13.4 $\sin i$ & 0.09 & 0.0 & None\\
& b & 541.0 $\sin i$ & 1.5 & 0.31 & \\
 \hline
\object{GJ~876} & d & 5.89 $\sin i$ & 0.021 & 0.0 & None\\
& c & 177.9 $\sin i$ & 0.13 & 0.27 & \\
 \hline\hline
\end{tabular}
\end{table*}

There are, however, many examples of hot Jupiters that have migrated
further inward than 0.1~AU. Orbits at $\sim$~0.05~AU are common and
some objects have been found as close as $\sim$~0.02~AU. It is less
likely that interior planets would survive in such systems. If we
assume that all remaining interior mass is accreted by the star
(which is not always true, as demonstrated below) then the total
surviving mass reduces to the exterior surviving mass which, as
illustrated in Fig.\ref{figure:7}, shows a clear trend with disk
maturity. The later the migration episode the more disk mass will
remain: from $\sim 30\%$ for our earliest scenario to ~$\sim 70\%$
for our latest. None of our simulations support the conjecture of a
near-complete loss of solids from the swept zone.

\subsection{The interior planets.}\label{interior}

In all scenarios, barring the latest, giant planet migration was
found to stimulate accretion within the portion of the disk
shepherded inward. By the end of the simulations, this mass had
accumulated into one or more massive planets within 0.1~AU. Details
of these planets and the giant, including their simulation ID, mass,
$a$, $e$, and the presence of resonances are given in Table
\ref{table:4}.

\begin{figure}
 \resizebox{\hsize}{!}{\includegraphics{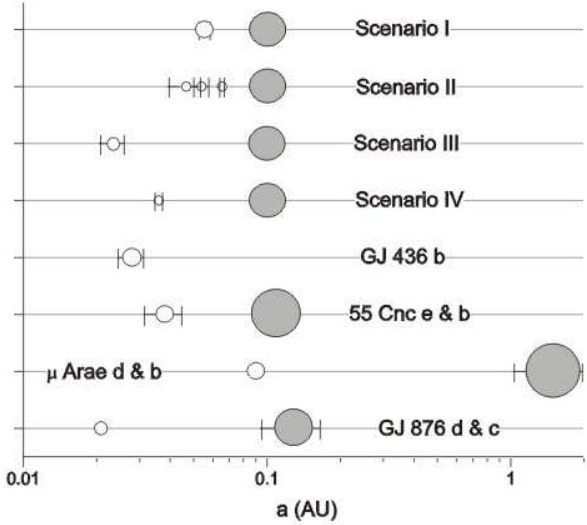}}
 \caption{Comparison of computer generated interior planets with the
 interior regions of four known "hot Neptune" systems.}
 \label{figure:8}
\end{figure}

A single interior planet was the most common result, but in one case
there were three survivors. Their masses ranged between $\sim
2-16~\mathrm{M}_{\oplus}$ with semi-major axes between $\sim$ 0.02
-- 0.07 AU. None of them remained in resonance with the giant even
though some of their precursor bodies would have been originally
been pushed inward at first order resonances (see
Fig.~\ref{figure:4}). These resonances were broken during the final
accretion phase within 0.1~AU. In Scenario II however, the three
planets that remain are all in resonant relation with each other.
The inner planet is in the 5:4 resonance with the middle planet and
the 5:3 resonance with the outer planet. The middle planet is in the
4:3 resonance with the outer planet, giving a 5:4:3 commensurability
overall. It is possible that such a relationship could act to
stabilize the orbits of these planets, but they are closely spaced,
the orbits of OLI9 and ICE9 cross, and the giant acts to perturb the
system, so we suspected accretion here to be incomplete. Running the
inner system of Scenario II for an additional 1.0 Myr resulted in a
prompt giant impact between the inner pair (ICE9 \& OLI9, see
Table~\ref{table:4}), followed by a longer phase of interaction of
the two 9.02 and 5.58~$\mathrm{M}_{\oplus}$ survivors. Their orbits
gradually became more elliptical, especially that of the lighter
outer planet (ICE1). Seven close encounters followed causing an
outward scattering and a further excitation of the outer planet's
eccentricity. Just before the million years was up, ICE1 encountered
the giant for the first time and 18 close encounters later they
collided. Thus, the final outcome for Scenario II was a single
remaining inner planet, separated from the giant by 12.9 mutual Hill
radii with a mass of 9.02~$\mathrm{M}_{\oplus}$, $a$ = 0.0465 AU,
$e$ = 0.132.

Generation of these massive interior planets by these simulations is
particularly interesting as three examples of short period
Neptune-mass objects have been recently discovered in the systems
\object{GJ~436} \citep{butler}, \object{55~Cancri} \citep{mcarthur}
and \object{$\mu$~Arae} \citep{santos2} and a short period planet
about half as massive may also have been detected in the
\object{GJ~876} system \citep{rivera}. These systems are compared
with those generated here in Table \ref{table:4} and
Fig.~\ref{figure:8}. In three of these natural systems the inner
Neptune-mass planet is accompanied by more than one outer giant and,
in the cases of 55 Cancri and GJ~876, the innermost giant is placed
close to where our simulated giant ends its migration. The results
of the simulations do have a particular resemblance to reality in
these two cases. The best matches are given by Scenario I, where the
mass and orbital radius of the interior planet are similar to the
$m\sin i$ and $a$ of 55~Cnc~e, and Scenario III where the
resemblance is closer to the configuration of GJ~876~d~\&~c. One
might speculate therefore that, rather than hot Neptune type planets
forming far out in the disk and migrating inward to their present
location, they might, as illustrated here, form at these locations
from disk material shepherded and compacted by a migrating giant.

\begin{figure*}
 \sidecaption
 \includegraphics[width=12cm]{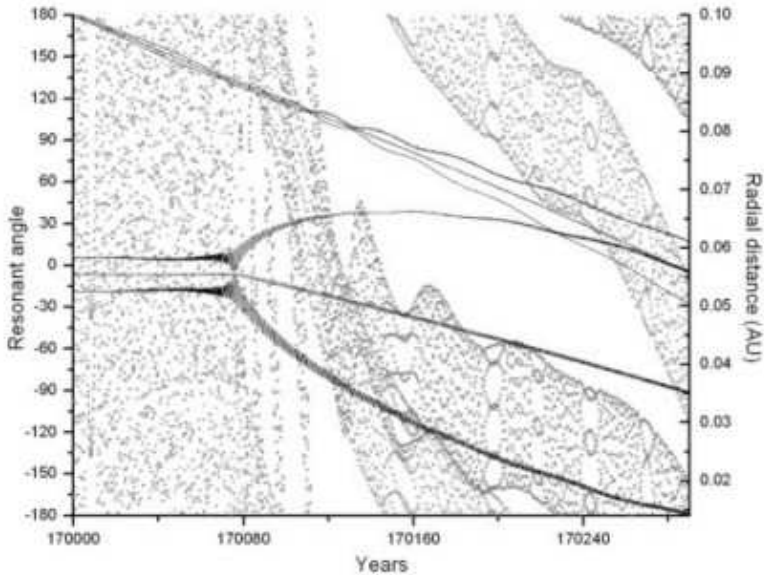}
 \caption{Scenario I: collision of the inner planet with the central
 star as the giant moves inward to 0.05 AU. Resonant angles for the
 2:1 resonance are read on the left hand axis: light grey symbols
 plot the angle $\phi_1 = 2\lambda'-\lambda-\varpi'$ and dark grey
 symbols plot the angle $\phi_2 = 2\lambda'-\lambda-\varpi$ where
 $\lambda',\varpi'$ and $\lambda,\varpi$ are the mean longitudes and
 longitudes of periastron for the outer and inner planet respectively.
 The semi major axis, periastron and apastron for both the giant and the
 inner planet are plotted as black lines and are read on the right
 hand axis.}
 \label{figure:9}
\end{figure*}

Against this proposition is the case of GJ~436 where the hot Neptune
appears unaccompanied by an exterior giant; although \citet{butler}
did detect a linear velocity trend in their data implying the
possible existence of a more distant companion. The primary of this
system is a $\sim 0.4~\mathrm{M}_{\odot}$ red dwarf star which might
affect the comparison. Giant planets appear to be rare in red dwarf
systems and it may be that they do not form efficiently from the
lower mass protoplanetary disks expected around low mass stars
\citep{laughlin1}. GJ~436~b may therefore be this system's largest
planet, rather than a secondary object, which could indeed have
formed at large radius before migrating in. However, not all red
dwarf stars lack giant planets. In the one known case where giants
are present (GJ~876) and past migration may have played a role in
their current configuration \citep[e.g.][]{snellgrove}, an interior
half-hot Neptune appears to be present. The hot Neptune in the
$\mu$~Arae system is accompanied by a giant, but it is situated much
further out (at 1.5~AU) than the final location of the giant in the
simulations. No simulations have yet been performed to evaluate the
outcome of stopping the giant migration at 1.5~AU, but at the point
in the runs where the giant passes through 1.5~AU (at $\sim$
140\,000 years) about 60\% of the original mass of the disk is
compacted within $\sim$ 1~AU. This is enough mass to assemble a $\ga
13~\mathrm{M}_{\oplus}$ hot Neptune over a longer period although,
since $\mu$~Arae~d and $\mu$~Arae~b are separated by $\sim$ 22
mutual Hill radii, one might expect some additional smaller planets
to have formed and perhaps to have survived to the present. Less
massive interior planets of $\sim 2-7~\mathrm{M}_{\oplus}$ are
produced in the simulations and may exist in nature as smaller
versions of the hot Neptunes already discovered. GJ~876~d may
represent the first discovery of a planet in this mass range.
However, such objects would only produce a $\sim 1 -
3~\mathrm{m~s}^{-1}$ stellar radial velocity at $\sin i = 1$, a
value that is close to the observation limit, so others will have
escaped detection so far.

Another reason for the apparent rarity of hot Nepture type objects
could be that many hot Jupiter systems are more compact than 0.1 AU;
a typical example being 51~Pegasi~b: $m \sin i \approx
0.5~\mathrm{M_J}$, $a \approx 0.05$~AU \citep{mayor}. For a planet
shepherded well within this distance, significant eccentricity
excitation by the giant companion may cause it to impact the star.

To examine this possibility, Scenarios I--IV were run for an extra
300 years, allowing the giant in each case to migrate further in to
stop at 0.05~AU. In Scenarios I \& II all the interior planets were
driven into the central star. The mechanism at work is illustrated
for the case of Scenario I in Fig.\ref{figure:9} where the semi
major axis, periastron and apastron distances, for both the giant
and terrestrial planet, and the resonant angles for the 2:1
resonance, are all plotted against time. The orbit of the interior
planet is initially undisturbed, but when the giant has moved inward
to 0.087 AU the planet is captured into the 2:1 resonance as can be
seen from the libration of the resonant angles. From this point the
planet is pushed in ahead of the giant at the 2:1 resonance, its
eccentricity increasing progressively. By the time the planet has
reached $a \approx 0.035$~AU, its eccentricity has increased to $e
\approx 0.6$ and impact with the star occurs at periastron. The
events in Scenario II were similar: the migrating giant in this case
caused the three interior planets present to accrete each other,
capturing the single 14.6~$\mathrm{M}_{\oplus}$ survivor into the
2:1 resonance. From here, evolution proceeded as in
Fig.\ref{figure:9}, the planet eventually hitting the star through
having been forced into a tighter, more elongated, orbit.

\begin{figure*}
 \sidecaption
 \includegraphics[width=12cm]{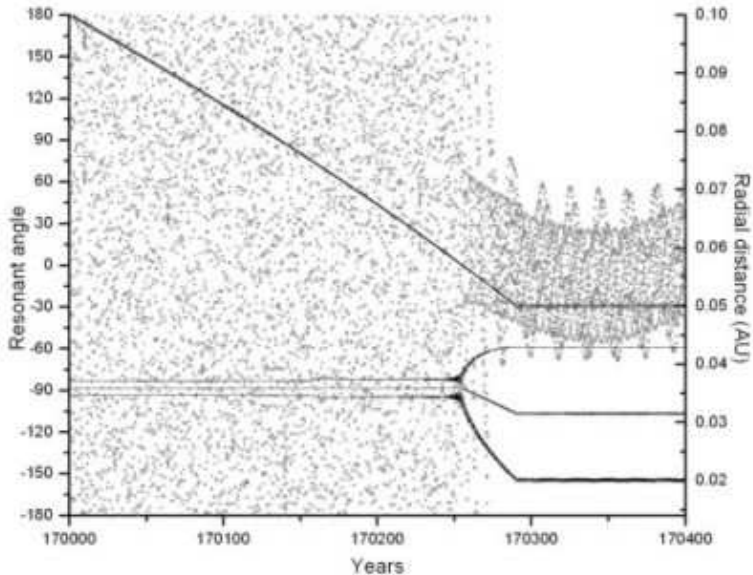}
 \caption{Scenario IV: survival of the inner planet as the giant moves
 inward to 0.05 AU. Resonant angles for the 2:1 resonance are read on
 the left hand axis: light grey symbols
 plot the angle $\phi_1 = 2\lambda'-\lambda-\varpi'$ and dark grey
 symbols plot the angle $\phi_2 = 2\lambda'-\lambda-\varpi$. The
 semi major axis, periastron and apastron for both the giant and the
 inner planet are plotted as black lines and are read on the right
 hand axis.}
 \label{figure:10}
\end{figure*}

In Scenarios III \& IV, both the interior planets survived this
additional migration by the giant. This is because their initial
orbits were closer to the star (see Table \ref{table:4}) so the
planet only becomes captured at the 2:1 resonance much later, or not
at all. The interior planet in Scenario III remained stable as the
2:1 resonance did not reach its location and sweeping higher order
resonances had no apparent effect. Survival of such a planet for the
long term is therefore probable, since orbital decay due to tidal
interaction with the star is expected to be very slow (see below).
In Scenario IV, the interior planet was captured into the 2:1
resonance late: at the point where the giant reached 0.057~AU. Its
orbit was compressed and elongated to $a$ = 0.032~AU and $e$ = 0.35
but the progressive increase in eccentricity ceased when the giant
finished migrating and impact with the star was avoided (see
Fig.\ref{figure:10}).

The effect of tidal interactions with a slowly rotating star will
cause the lower mass planet obtained in Scenarios III and IV to
migrate inward slowly. For a circular orbit tidal dissipation occurs
within the star only, with turbulent convection in the stellar
envelope being responsible for dissipating the tidally induced
motions. The estimated orbital evolution time in this case is given
by \citep{terquem}:
\begin{equation}
\label{tides}
t_{orb} \simeq 2.8 \times 10^{-4} \left(\frac{M_{\odot}}{m_p} \right)
\left(\frac{P}{1 \; {\rm day}} \right)^{\frac{13}{3}} \; {\rm Gyr}.
\end{equation}

\noindent where $P$ is the orbital period and $m_p$ is the mass of
the planet. For a 15 M$_{\oplus}$ planet with an orbital period of
1.5 days this gives a orbital decay time of $\sim 36$ Gyr -- i.e.
comfortably longer than the age of the universe. For a planet on an
eccentric orbit, tidal dissipation within the planet becomes
important due to more effective dissipation within the solid body
\citep[e.g.][]{goldreich1}. The raising and dissipation of tides
within the planet leads to eccentricity damping, and the low moment
of inertia of the planet ensures that it maintains near-synchronous
rotation. This also means that effective removal of orbital angular
momentum can only be achieved through tidal dissipation in the star.
We therefore expect that a planetary system consisting of an
exterior gas giant planet and an interior rocky planet, orbiting in
close proximity to a solar type star, will evolve such that the
rocky planet orbit decays on a time scale on the order of that given
by Eq.~(\ref{tides}). The inner planet will have an eccentricity
determined by a balance between tidal dissipation originating in the
planet itself and eccentricity excitation caused by the exterior
giant. In general, the long term effects of this will be to make the
orbits of both the inner terrestrial and outer giant planet more
circular. This will be enhanced by the tidal interaction between the
gas giant and the central star which will also tend to circularize
its orbit \citep[e.g.][]{rasio}, though we note that this effect is
not particularly relevant for our simulations as the giant planet is
assumed to maintain a near-circular orbit through tidal interaction
with the gas disk. For the specific case of Scenario IV, where the
inner planet is in 2:1 resonance with the gas giant, tidal
dissipation will cause the rocky planet orbit to circularize at a
smaller semi major axis, probably removing the inner planet from the
2:1 resonance in the process. Using Eq.~(\ref{tides}) we estimate
that the rocky planet formed in Scenario III will spiral into its
host star on a time scale of $\sim 54$ Gyr, and that formed in
Scenario IV on a time scale of $\la 700$ Gyr, again both comfortably
longer than the age of the universe.

\begin{table*}
\caption{External surviving protoplanets at 170\,000 years.} %
\label{table:5}  %
\centering  %
\begin{tabular}{c| c c c c c c c c c c c}
 \hline\hline %
Scenario & N & $\overline{m}_{\mathrm{proto}}~(\mathrm{M}_{\oplus})$
& $m_{\mathrm{max}}~(\mathrm{M}_{\oplus})$ &
$\overline{a}~(\mathrm{AU})$ & $a_{\mathrm{min}}~(\mathrm{AU})$ &
$a_{\mathrm{max}}~(\mathrm{AU})$ & $\overline{e}$ &
$e_{\mathrm{min}}$ & $e_{\mathrm{max}}$ &
$\overline{i}~\degr$ & $i_{\mathrm{max}}~\degr$\\
 \hline
I & 32 & 0.13 & 0.86 & 4.59 & 0.63 & 9.31 & 0.48 & 0.19 & 0.81 &
5.65 & 25.18\\
II & 32 & 0.13 & 0.86 & 7.04 & 1.31 & 28.24 & 0.53 & 0.24 & 0.86 &
3.64 & 9.35\\
III & 25 & 0.19 & 0.67 & 5.81 & 1.84 & 13.07 & 0.47 & 0.16 & 0.87
& 4.33 & 14.90\\
IV & 23 & 0.42 & 0.99 & 8.18 & 1.21 & 33.5 & 0.53 & 0.13 & 0.89 &
2.71 & 11.54\\
V & 16 & 0.55 & 1.34 & 10.26 & 2.42 & 40.97 & 0.52 & 0.10 & 0.96 &
4.62 & 11.33\\
 \hline\hline
\end{tabular}
\end{table*}

If the disk shepherding and compaction scenario advanced here has
some validity, then it suggests the formation of hot Neptunes or
lesser massive terrestrial planets as a by-product of giant planet
migration. However, if the giant comes to rest too close to the star
($\lesssim$ 0.05 AU) then the effects of mean motion resonances may
cause interior planets to hit the star. The best prospect
therefore of detecting close orbiting terrestrial type planets of
$\gtrsim\mathrm{M}_{\oplus}$ might be in systems with a
circumscribing giant at $a \sim$ 0.05 -- 1.5 AU.

\subsection{The exterior scattered disk.}\label{exterior}
As the giant migrated through the inner disk it scattered $\sim$
30--70\% of the disk mass into external orbits (see
Figs.~\ref{figure:5}~\&~\ref{figure:7} and Table \ref{table:3}). In
each case, a diffuse and dynamically excited external disk was
generated, composed predominantly of protoplanet material
($f_{\mathrm{proto}} \approx 0.7 - 0.9$): individual protoplanets
having a wide range of mass and orbital parameters. Data for the
external protoplanets are presented in Table \ref{table:5}, giving
their number, mean and maximum masses and orbital inclinations, and
their mean, minimum and maximum semi major axes and eccentricities.
It can be seen from the data that the number of external
protoplanets reduced and their masses increased with disk maturity.
Similarly, there is a tendency for protoplanets from a more mature
disk to be more widely scattered. The former trend is primarily due
to previous accretion, whereas the latter is a result of the
decreased dynamical friction operating at later times.

The ejecta that comprised the scattered disk were spread over a much
wider volume than their original location at $<$~4~AU. Planetesimal
orbits were damped quite rapidly by gas drag to form a thin ($\sim 1
- 2~\mathrm{M}_{\oplus}$) external disk with a surface density $\la$
a few percent of the pre-existing $\Sigma_{\mathrm{s}}$.
Protoplanets were often in highly eccentric orbits, passing well
beyond the confines of the original disk but with their periastra
still located close to the location of their scattering within 4~AU.
Mean orbital inclinations were comparable to the solar system
planets, but with a larger number of outliers as high as $i \approx
25\degr$ (see Fig.\ref{figure:5}). Thus, by selectively pushing
planetesimals inward and widely scattering its external ejecta, the
migrating giant partially evacuates a cavity within its swept zone.

Further accretion in this disk will therefore be characterized by
low $\Sigma_{\mathrm{s}}$ and high random velocities, reducing both
the mass available and the effect of gravitational focussing. In
some collisions, impact velocities could be high enough to cause
disruption of the protoplanets rather than accretion and a reversal
of growth \citep{agnor}. Long evolution times are implied for the
mass contained in the scattered disk to rearrange itself into a
smaller number of planets in stable orbits. This final configuration
cannot be predicted from the juvenile stage illustrated in
Fig.\ref{figure:5}.

No original matter more distant than the giant's starting position
was modeled here. Objects in the external scattered disk traverse
more widely than this and so could interact with matter in the outer
disk beyond $\sim$ 6~AU. A fresh supply of planetesimals would be
encountered which could act to circularize and contract protoplanet
orbits via dynamical friction \citep{thommes1}. Alternatively,
encounters with other giant planets remaining in the outer system
could have a role to play in clearing material via ejection. The
long term end product of the mass scattered by the migrating giant
therefore depends partially on the nature of the outer disk: what
other planets have formed there and its remaining population of
small bodies. In a hot Jupiter system where no other gas giants have
formed, we might speculate over the resulting planetary
configuration at $\sim 1$~Gyr: a hot Neptune, or lesser massive
terrestrial planet at $\sim 0.05$~AU, the giant at $\sim 0.1$~AU,
then from $\sim 0.5$~AU a succession of Mars to Earth-mass planets,
some still in eccentric or inclined orbits, extending as far out as
$\sim 5$~AU. Beyond this, the content of the original outer disk
would determine what is to be found.

The ejection of low mass planets to large ($\sim$30~AU) semi-major
axes may have some influence on the observed morphology of the
system in its debris disk phase. Systems such as \object{Fomalhaut}
and \object{$\varepsilon$~Eridani} \citep{wyatt,quillen} have
observational features that have been explained by resonant trapping
of planetesimals \citep{wyatt} or dust grains in mean motion
resonances with a planet. We suspect that the dust trapping proposed
by \citet{quillen} will also occur for lower mass planets (they
considered $m_{planet} \approx 30 \mathrm{M}_\oplus$), but is likely
to occur for resonances that lie closer to the planet than the 3:2
resonance that was dominant in their study.

\begin{table}
\caption{Total mass and number of protoplanets orbiting within, or
crossing, the habitable zone (0.84 -- 1.67~AU) at the end of the
simulation.} %
\label{table:6}  %
\centering  %
\begin{tabular}{c| c c c}
 \hline\hline
Scenario & In & Crossing & Total\\
 \hline\hline
I & 0.18~$\mathrm{M}_{\oplus}$ & 2.13~$\mathrm{M}_{\oplus}$ &
2.31~$\mathrm{M}_{\oplus}$\\
& N = 1 & N = 8 & N = 9\\
 \hline
II & 0.18~$\mathrm{M}_{\oplus}$ & 1.79~$\mathrm{M}_{\oplus}$ &
1.97~$\mathrm{M}_{\oplus}$\\
& N = 1 & N = 5 & N = 6\\
 \hline
III & & 1.28~$\mathrm{M}_{\oplus}$ & 1.28~$\mathrm{M}_{\oplus}$\\
& & N = 4 & N = 4\\
 \hline
IV & 1.17~$\mathrm{M}_{\oplus}$ & 2.00~$\mathrm{M}_{\oplus}$ &
3.17~$\mathrm{M}_{\oplus}$\\
& N = 2 & N = 3 & N = 5\\
 \hline
V & & 1.34~$\mathrm{M}_{\oplus}$ & 1.34~$\mathrm{M}_{\oplus}$\\
& & N = 1 & N = 1\\
 \hline\hline
\end{tabular}
\end{table}

\begin{figure}
 \resizebox{\hsize}{!}{\includegraphics{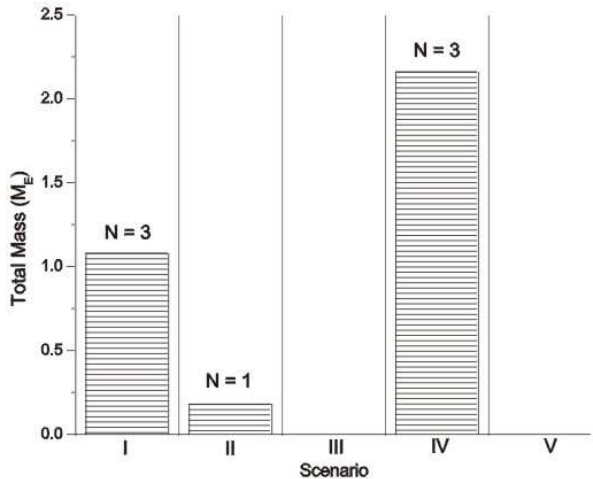}}
 \caption{Number and total mass of protoplanets predicted in the
 habitable zone assuming orbital re-circularization with
 conservation of angular momentum.}
 \label{figure:11}
\end{figure}

What of the probability of one of these surviving planets residing
in the system's habitable zone? Habitable zones are dynamically
stable in hot Jupiter systems \citep{menou} so a planet forming
there would have a stable orbit if sufficiently well-spaced from
neighbors. The simulations presented here cannot provide any
numerical estimate that might address this question: the simulations
have not been run for long enough and, in any case, ignore the
strong potential influence of an outer disk. However, protoplanets
are found within, or with their orbits passing though, the habitable
zone \citep[taken to be 0.84 -- 1.67~AU;][]{kasting} at the end of
each run (see Table \ref{table:6}). Thus, some mass is available for
forming a completed planet in the right place, especially in view of
the fact that damping of the protoplanet's orbits from interaction
with outer disk material, collisional damping during ensuing
accretion, destructive collisions, and gas drag on remaining
planetesimals and debris could act to return some material into
closer orbits. Assuming eventual re-circularization of orbits with
conservation of angular momentum, the data in Table \ref{table:6}
convert to those illustrated in Fig.\ref{figure:11} which shows, for
each scenario, the number of protoplanets and their total mass
predicted in the habitable zone. Protoplanets occupy the habitable
zone in three out of the five cases and, in two scenarios, more than
an Earth-mass of material is present. Thus, since there is dynamical
room available within the habitable zone of a hot Jupiter system, it
is perhaps more likely than not that long term evolution of the
scattered external disk could result in a terrestrial planet being
located there.

\section{Caveats and future model improvements}\label{caveats}
Inevitably a number of assumptions have been used in constructing the models
presented in this paper, with the consequence that potentially
important physical processes have been omitted. Here we discuss the
possible implications of these for our results, whilst noting that work is
underway to include them in future models: \\
({\it i}) {\em Gap formation \& cavity clearing:} \hspace{1mm}
A giant planet is expected to form an annular gap in the gas disk
centred around its orbit. As a Jupiter mass planet migrates inward, hydrodynamic
simulations indicate that the inner disk becomes depleted of gas, forming
a low density cavity there \citep[e.g.][]{nelson1}. This arises in part
because the viscous time scale in the inner disk is shorter than the
migration time. The depletion is probably enhanced artificially by the use of
an outflow boundary condition at the disk inner edge, which for computational
reasons is located at a radius much larger than the expected surface of the
central star. At this time it is not known accurately
to what degree the inner disk is
depleted. We have assumed an undepleted disk, and the effects of
this will be the subject of future investigation.
We note that our use of a 0.5 M$_J$ planet will lead to significantly less
gas depletion than obtained in hydrodynamic
simulations performed using more massive planets. \\
The effect of depleting the gas disk
interior to the giant planet will be to reduce the gas drag experienced
by the planetesimals. Reducing the dissipation level is likely to
lead to greater scattering of bodies into the outer disk by the giant planet,
which may reduce the efficiency of planet formation in the inner disk.
We note, however, that the loss of planetesimals into the central star due
to gas drag will also diminish. Models similar to those
presented here, but coupled to an evolving gas disk model, will be
the subject of a future publication. \\
({\it ii}) {\em Gas disk removal:} \hspace{1mm} Implicit within our models
is a mechansim for preventing the
gas giant planet from migrating all the way into the central star.
While the presence of a magnetospheric cavity has been cited as
a possible stopping mechanism for hot Jupiters \citep{lin1},
we note that planets
are found to exist with a range of semi major axes for which this mechanism
cannot be invoked. A more plausible reason for migration halting
is the removal of the gas disk during migration. The removal of the outer gas
disk would remove a source of dissipation for the scattered protoplanets
and planetesimals, increasing the accretion time scale
for any planets forming exterior to the giant. \\
({\it iii}) {\em Giant planet eccentricity evolution.} \hspace{1mm}
Our model assumes that type II migration is associated with an
eccentricity damping which maintains a near-circular orbit. However
the question of whether a giant planet has its eccentricity damped
or excited by the protoplanetary disk remains a point of ongoing
debate \citep{papaloizou2,goldreich2,ogilvie}. In the absence of
eccentricity damping, interaction of the giant with the solids disk
may cause modest excitation, but not at a level to significantly
effect our results. If the gas disk was to drive significant
eccentricity increase then this would probably lead to a stronger
interaction between the giant and the solids disk resulting in
greater scattering and a more efficient clearing of material.\\
({\it iv}) {\em Planetesimals exterior to the gas giant:}
\hspace{1mm} There will be a population of planetesimals exterior to
the giant planet whose size distribution is unknown. Planetesimals
with radii $< 1$ km will migrate in behind the gas giant due to gas
drag, providing a source of material that may be accreted by
outwardly scattered protoplanets that reside interior to 5~AU, and
which may
contribute to the damping of their inclination and eccentricity. \\
({\it v}) {\em Planetesimal size evolution:} \hspace{1mm} The model
we have used assumes that a population of 10 km sized planetesimals
coexists with a population of larger protoplanets at the beginning
of gas giant migration. Collisions between planetesimals may lead to
both their growth and fragmentation, with a range of sizes
developing. At present these processes are not included in our
model, and their effect on the outcome of our calculations is
unclear. We simply note that smaller bodies produced by
fragmentation will experience larger gas drag forces, leading to a
more rapid in--spiral toward the central star, and reduced
eccentricity that assists rapid accretion by the protoplanets. The
build--up of larger bodies will lead to reduced effectiveness of gas
drag, and a greater probability of scattering. \\
({\it vi}) {\em Type I migration:} \hspace{2mm} We have neglected
the effects of type I migration which operates for non--gap forming
sub--Jovian planets (Ward 1997). This may become important for
bodies more massive than $\sim 1$ M$_{\oplus}$, causing inward
migration and strong eccentricity damping. For our model disk
parameters type I migration proceeds faster than the gas giant
migration for bodies more massive than $\sim 5$ M$_{\oplus}$. We
note that such bodies usually form during a rapid burst of accretion
as the planet approaches the star, so their formation is unlikely to
be greatly affected by type I migration. Their subsequent evolution
will depend on the gas density in the inner disk, which at present
is unknown, as discussed in point ({\it i}) above. \\
We further comment that a consistent model of giant planet formation
{\em via} the core-instability model is difficult to achieve if type
I migration operates as efficiently as current calculations suggest.
Our model of a gas--giant forming and then slowly migrating inward
implicitly contains the assumption that type I migration does not
operate efficiently. We note that recent simulations of low mass
planets in turbulent disks suggest that some low mass planets may
avoid rapid inward migration \citep[e.g.][]{nelson2,nelson3}, so the
role played by type I migration during planet
formation is unclear at present. \\
Type I disk-planet interactions are also thought to include
significant eccentricity damping on a timescale $\sim$100 times
shorter than the migration timescale \citep{papaloizou1}, an effect
that could be relevant to the picture we have presented. The
influence of this in simulations would be to exert additional
dissipation on the protoplanets and hence increase their probability
of being shepherded by the giant. Such behavior would have its
greatest impact in more evolved systems where dissipation is weaker,
such as our Scenarios IV and V. Type I damping would also effect the
evolution of the scattered external disc and would be an additional
influence acting to return some planet-forming material into closer
orbits. Its integrated effect however would depend on how much of
the nebula's gas remains and how much longer it lasts for.
Eccentricity damping from a thin residual gas disk could act to
circularize the orbits of any final planetary configuration
\citep{kominami}, whereas if ample gas persists and type I migration
is also in force, a pileup of mass could also occur exterior to the
giant as migrating protoplanets are entrained in its exterior
mean-motion resonances \citep{thommes3}. It is possible therefore
that a modest degree of type I eccentricity damping could promote,
rather than hinder, the accretion of a planetary system external to
the orbit of the giant planet. \\

\section{Conclusions.}\label{conclusions}
Whilst this study has addressed the effect of giant planet migration
through inner disks of varying maturity, we have not examined the
effect of changing some of the basic parameters that could influence
accretion, shepherding and scattering behavior. These might include
varying the giant's mass, its migration time ($\tau_{\nu}$) and the
mass of the nebula ($f_{\mathrm{neb}}$). Such a study will be the
subject of future work, which will also address some of the issues
raised in Section~\ref{caveats} concerning extensions to the basic
physical model. It is likely that varying the mass of the nebula
would result in a change in accretion time scales and a
corresponding increase or decrease in the mass of the surviving
planets, but not in their elimination entirely. It would also affect
the level of dissipation in the disk which influences the dynamical
partitioning of disk mass brought about by the migration. Increasing
the mass of the giant would raise the overall effect of scattering,
as would extending the migration time which increases opportunities
for repeated and incremental scatterings \citep{mandell}. It seems
reasonable to speculate that varying any of these parameters within
reasonable limits would still produce a result that, whilst varying
in detail, remains consistent with our established picture.

There are three principal conclusions that arise from this work.
\begin{enumerate}
\item Migration of a giant planet through an inner disk partitions
the mass of that disk into internal and external remnants. The
fraction of disk mass in either remnant is dependent on the level of
dissipation and is thus sensitive to the maturity of the disk
material at the time of the migration episode. Late migration favors
the escape of more material into external orbits. The survival of an
inner remnant is also sensitive to the final position of the giant
at the end of migration, this becoming increasingly unlikely in hot
Jupiter systems with $a \la$~0.05~AU. The concept that giant planet
migration would eliminate all the mass in its swept zone is not
supported by our results.
\item Hot Neptunes and lesser massive terrestrial planets are a
possible by-product of type II migration, being formed from an inner
system disk compacted by a migrating giant. Future searches of hot
Jupiter systems for radial velocity signals close to the current
detection limit might uncover more examples of these planets.
\item Our results are supportive of the eventual accumulation of a
number of terrestrial planets orbiting exterior to the giant,
including within the system's habitable zone. Thus, the early
evolution and the final architecture
of Hot Jupiter systems does necessarily eliminate their possibility
of hosting Earth-like planets.
\end{enumerate}


\bibliographystyle{aa}

\listofobjects

\end{document}